\documentclass[aps,pre,preprint,groupedaddress]{revtex4-1}
\usepackage{graphicx}
\begin{document}

% Use the \preprint command to place your local institutional report
% number in the upper righthand corner of the title page in preprint mode.
% Multiple \preprint commands are allowed.
% Use the 'preprintnumbers' class option to override journal defaults
% to display numbers if necessary
%\preprint{}

%Title of paper
%%%%%%%%%%%%%%\title{}
\title{ Instability of group formation through indirect reciprocity
under imperfect information and implementation error} 
% repeat the \author .. \affiliation  etc. as needed
% \email, \thanks, \homepage, \altaffiliation all apply to the current
% author. Explanatory text should go in the []'s, actual e-mail
% address or url should go in the {}'s for \email and \homepage.
% Please use the appropriate macro foreach each type of information

% \affiliation command applies to all authors since the last
% \affiliation command. The \affiliation command should follow the
% other information
% \affiliation can be followed by \email, \homepage, \thanks as well.
\author{A. Isagozawa}%
\affiliation{%
Faculty of Humanities and Social Sciences,%
Iwate University, Morioka 020-8550, Japan%
}%
\author{T. Shirakura}%
\email[]{shira@iwate-u.ac.jp}%
\affiliation{%
Faculty of Humanities and Social Sciences,%
Iwate University, Morioka 020-8550, Japan%
}%
\author{S. Tanabe}%
\affiliation{%
Faculty of Humanities and Social Sciences,%
Iwate University, Morioka 020-8550, Japan%
}%

%%%%%%%%%%%%%%%%%\author{}
%\email[]{Your e-mail address}
%\homepage[]{Your web page}
%\thanks{}
%\altaffiliation{}
%%%%%%%%%%%%%%%%%%%\affiliation{}

%Collaboration name if desired (requires use of superscriptaddress
%option in \documentclass). \noaffiliation is required (may also be
%used with the \author command).
%\collaboration can be followed by \email, \homepage, \thanks as well.
%\collaboration{}
%\noaffiliation

\date{\today}

\begin{abstract}

	Indirect reciprocity is a key mechanism behind 
the evolution of cooperation.
Oishi et al. analytically showed 
the formation of two exclusive
groups under the KANDORI assessment rule
in the case of perfect information and no implementation
error, regardless of the population size $N$.
Here, we numerically show the formation of many exclusive groups
under the JUDGING assessment rule in the same case.
Introducing degrees of exclusive groups, 
 we numerically examine
the stability of the group formation under imperfect information
and implementation error. 

\end{abstract}


\maketitle

%%%%%%%%%%%%%%%%%%%%%%%%%%%%%%%%%%%%%%%%%%%%%%%%%%%%%%%%%%%%%%%%%%%%%%%%%%%%%%%%%

\section{Introduction}

	Indirect reciprocity is a key mechanism driving the evolution of 
cooperation.\cite{Nowak} 
One feature of indirect reciprocity is that helpful acts are returned,
not by the recipient as in direct reciprocation,
but by third parties:
 players collect
information about other player's behavior, and
 determine their actions by using this information. 
This behavior requires the following two
modules: (a) an assessment rule for the acts of others
as good or bad, and (b) an action rule specifying how to act toward others
based on that assessment.

	Assessment rules are classified into three types, depending on 
what information they use. A first-order assessment rule only takes
into account whether a donor $X$ helps a recipient $Y$.
A second-order assessment rule takes also into account the image
of the recipient $Y$. A third-order assessment rule additionally
takes into account the image of the donor $X$.

	Assuming binary assessments,
Ohtsuki and Iwasa \cite{Ohtsuki1, Ohtsuki2} showed 
that among the resulting possible strategies, only eight
lead to a stable regime of mutual cooperation under public
and perfect information about others. 
These are called the leading eight.

	On the other hand,
Uchida \cite{Uchida} examined the effects of private and imperfect information
as well as those of implementation errors. 
Although Uchida concluded that private information leads to the collapse of
the sterner (KANDORI\cite{Kandori}) assessment rule, the results presented here
are somewhat different. 

	Oishi et. al.\cite{Oishi} showed the emergence of two exclusive
groups in the JUDGING assessment rule when information is 
perfect and private. In this study, we numerically examine the stability of the
exclusive groups under imperfect information and implementation errors.
Furthermore, we numerically show the emergence of many exclusive groups
in the third-order (JUDGING) assessment rule.

%%%%%%%%%%%%%%%%%%%%%%%%%%%%%%%%%%%%%%%%%%%%%%%%%%%%%%%%%%%%%%%%%%%%%%%%

\section{Model and Methods}

	We consider the donation game defined as follows:
there are $N$ players, each with their own opinion as to whether each of
the other players is good ($G$) or bad ($B$) (image matrix).
$\beta_{ij}(t)\in{G, B}$ represents player $i$'s opinion of player $j$
at a round $t$. Each element of the image matrix at the initial round
$t=1$, is $G$ with probability $p$ or $B$ with probability $1-p$.
The probability $p$ is called the initial trust probability and $0<p<1$.

	The game is repeated over a large number of rounds. In each round,
one player is randomly chosen as a donor and another 
 as a recipient. It is the donor' choice whether to cooperate or 
defect. We assume the decision follows an action rule.
If the donor cooperates, the payoff of the donor, $-c$, is less than 0
 and that of
the recipient, $b$, is greater than 0. If the donor defects, the payoffs of both players
are $0$. We assume $b \geq c$.

	All players observe which player is the donor, which is the recipient, 
and what 
the donor does to the recipient. Then, all players independently revise
their own opinion on the donor based on the observation and their
assessment rule.

	Here we use only one action rule, namely that the donor only
cooperates with recipients whom the donor regards as good.
On the other hand, we mainly study two assessment rules, which are called the
KANDORI and JUDGING ones. In Table 1, we show definitions of the assessment rules
used in this paper. 
  
%%%%%%%%%%%%%%%%%%%%%%%%%%%%%%%%%%%%%%%%%%%%%%%%%%%%%%
\begin{table}[b]
\caption{\label{tab:table1}
 Assessment of the donor as G (good) or B (bad) by observers after the donor 
cooperates with (C), or defects from (D) the recipient. Observers have pre-assessments
 of 
the donor and the recipient as to whether they are good or bad.
}
\begin{ruledtabular}
\begin{tabular}{|l|c|c|c|c|c|c|c|c|} \hline
pre-assessment to the donor & \multicolumn{4}{c|}{good} & \multicolumn{4}{c|}{bad} \\ 
\hline
pre-assessment to the recipient & \multicolumn{2}{c|}{good} & \multicolumn{2}{c|}{bad} &
\multicolumn{2}{c|}{good} & \multicolumn{2}{c|}{bad} \\
\hline
cooperate(C) or defect(D) & C & D & C & D & C & D & C & D \\
\hline
All C & G & G & G & G & G & G & G & G \\ 
All D & B & B & B & B & B & B & B & B \\
SCORING & G & B & G & B & G & B & G & B \\ 
TYPE 1 & G & B & G & G & G & B & G & G \\
KANDORI & G & B & B & G & G & B & B & G \\
STANDING & G & B & G & G & G & B & G & B \\
JUDGING & G & B & B & G & G & B & B & B \\
\end{tabular}
\end{ruledtabular}
\end{table}
%%%%%%%%%%%%%%%%%%%%%%%%%%%%%%%%%%%%%%%%%%%%%%%%%%%%%%%%%%%%%%%%%%%%%
  
  We consider two types of noises, following Uchida's paper.\cite{Uchida}
We investigate the effect of imperfect information, in which each 
 interaction is only observed by a fraction, $q<1$, of the population.
We also study the effect of implementation error $\epsilon$, which
is the probability that an intended help is not actually given.

%%%%%%%%%%%%%%%%%%%%%%%%%%%%%%%%%%%%%%%%%%%%%%%%%%%%%%%%%%%%%%%%%%%%%%%%%%%%%%%%
%%%%%%%%%%%%%%%%%%%%%%%%%%%%%%%%%%%%%%%%%%%%%%%%%%%%%%%%%%%%%%%%%%%%%%%%%%%%%%%%

\section{Results using the KANDORI assessment rule} 
	First, we show our results in the case of the KANDORI assessment rule.

\subsection{The case of $q=1$ and $\epsilon=0$}
	We first consider the case of perfect infomation, $q=1$, and no 
implementation error, $\epsilon=0$.  Oishi et al.\cite{Oishi}
 analytically  investigated this case and showed the formation of
two exclusive groups with ratios of $p$ and $1-p$, irrespective of the
population size, $N$.

\subsection{The case of $q<1$ and $\epsilon=0$}
   In this case, the results depend upon the initial state of the image matrix,
$\beta_{ij}(t=1)$, and the number of players, $N$.
We consider two initial states of an image matrix for some $p$ ($0<p<1$),
namely 
1) a random initial state that is randomly set to $G$ with a probability $p$ 
or $B$ with
a probabilty $1-p$, and 2) a uniform initial state where 
two exclusive groups with ratios $p$ and $1-p$ are initially formed.
In the latter case, we easily confirm that imperfection of the information,
$1-q$, does not destroy the two exclusive groups.

   In the former case, we can see that, after 
a large number of rounds and sorting (see Appendix A),
 two imperfect exclusive
groups still appear for small $N$ and small $1-q$, as seen in Fig. 1.
Here, we introduce the degree of the two exclusive groups, $DEG1$, as follows:
\begin{eqnarray}
DEG1 = \frac{1}{N^2} \sum _{j=1}^{N} \sum _{i=1}^{N} sign(\beta_{1j})\beta_{1i}
\beta_{ji}. 
\end{eqnarray}
$DEG1$ is equal to one when the two exclusive groups are perfectly formed, and becomes
zero when they are destroyed. In Fig. 2, the results of $DEG1$ versus $q$ are
shown for $N=100, 200$, and $400$. For larger numbers of players, 
we can see that smaller values of the
incompleteness of the information, $1-q$, disturbs the formation of the two 
exclusive groups from
a random initial state.

%%%%%%%%%%%%%%%%%%%%%%%%%%%%%%%%%%%%%%%%%%%%%%%%%%%%%
\begin{figure}[tbhp]
\begin{center}
\includegraphics{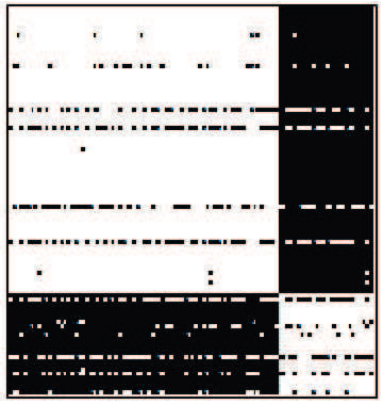}\\
\end{center}
\vspace{-0.4cm}
\caption{
Image matrix $\beta_{ij}$ at the 20,000th round
starting with a random initial state, where $N$=100,   
$q$=0.99, $p$=0.8, and $\epsilon=0$.
}
\end{figure}
%%%%%%%%%%%%%%%%%%%%%%%%%%%%%%%%%%%%%%%%%%%%%%%%%%%%% 

%%%%%%%%%%%%%%%%%%%%%%%%%%%%%%%%%%%%%%%%%%%%%%%%%%%%%
\begin{figure}[tbhp]
\begin{center}
\includegraphics{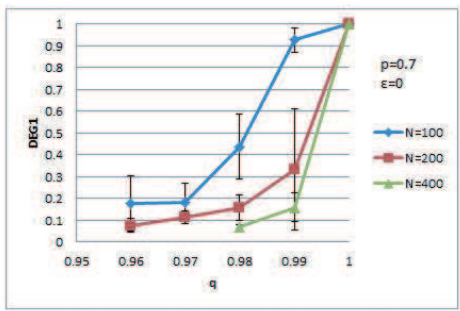}\\
\end{center}
\vspace{-0.4cm}
\caption{
$DEG1$ versus $q$ for $N=100, 200$ and 400
after a large number of rounds from a random initial state, where
$p=0.7$ and $\epsilon=0$ and the error bars show a standard
deviation of 10 samples with different random numbers.
}
\end{figure}
%%%%%%%%%%%%%%%%%%%%%%%%%%%%%%%%%%%%%%%%%%%%%%%%%%%%% 

\subsection{The case of $q=1$ and $\epsilon>0$}
	After a large number of rounds, two groups appear.
Players within the same group have the same opinions; 
those in different groups have opposite opinions.
The two groups are not necessarily exclusive.
In Fig. 3, we show that $DEG1$ linearly decreases
 with $\epsilon$,
irrespective of the population size $N$.

%%%%%%%%%%%%%%%%%%%%%%%%%%%%%%%%%%%%%%%%%%%%%%%%%%%%%
\begin{figure}[tbhp]
\begin{center}
\includegraphics{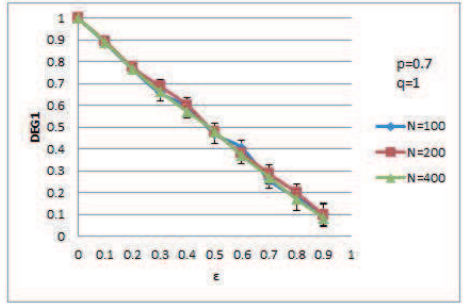}\\
\end{center}
\vspace{-0.4cm}
\caption{
 $DEG1$ versus $\epsilon$ for $N$=100, 200, and 400, where
$p$=0.7 and $q=1$; the error bars show the standard
deviation from 10 samples.
}
\end{figure}
%%%%%%%%%%%%%%%%%%%%%%%%%%%%%%%%%%%%%%%%%%%%%%%%%%%%% 

\subsection{The case of $q<1$ and $\epsilon>0$}
	In contrast with the case of $q<1$ and $\epsilon=0$,
there is no dependence upon different initial states with the same $p$
after a large
number of rounds. In Fig. 4, the results of $DEG1$ versus $q$ are
shown for $N=100, 200$, and $400$.
We obtain results similar to the case where $q<1$ and $\epsilon=0$
for a random initial state.

%%%%%%%%%%%%%%%%%%%%%%%%%%%%%%%%%%%%%%%%%%%%%%%%%%%%%
\begin{figure}[tbhp]
\begin{center}
\includegraphics{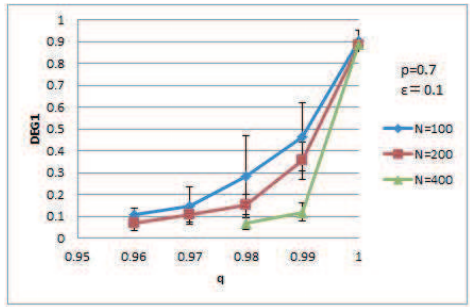}\\
\end{center}
\vspace{-0.4cm}
\caption{
$DEG1$ versus $q$ for $N$=100, 200, and 400
after a large number of rounds, where
$p=0.7$ and $\epsilon=0.1$; the error bars show the standard
deviation of 10 samples.
}
\end{figure}
%%%%%%%%%%%%%%%%%%%%%%%%%%%%%%%%%%%%%%%%%%%%%%%%%%%%%

\section{Results using the JUDGING assessment rule} 

We show results using the JUDGING assessment rule.

\subsection{The case of $q=1$ and $\epsilon=0$}

    Many exclusive groups are formed 
after a large number of rounds from a random initial state at some $p$, where
$0<p<1$, as shown in Fig. 5.
The sizes of these groups and their frequencies depend upon $p$. 
Fig. 6 shows relations between group sizes and frequencies. 
The larger the value of $p$, the greater the ease with which large groups are formed.
 When $p$ is small, it shows a behavior close to a power-low distribution. 
This distribution is almost the same for different population sizes $N$,
as shown in Fig. 7.

%%%%%%%%%%%%%%%%%%%%%%%%%%%%%%%%%%%%%%%%%%%%%%%%%%%%%
\begin{figure}[tbhp]
\begin{center}
\includegraphics{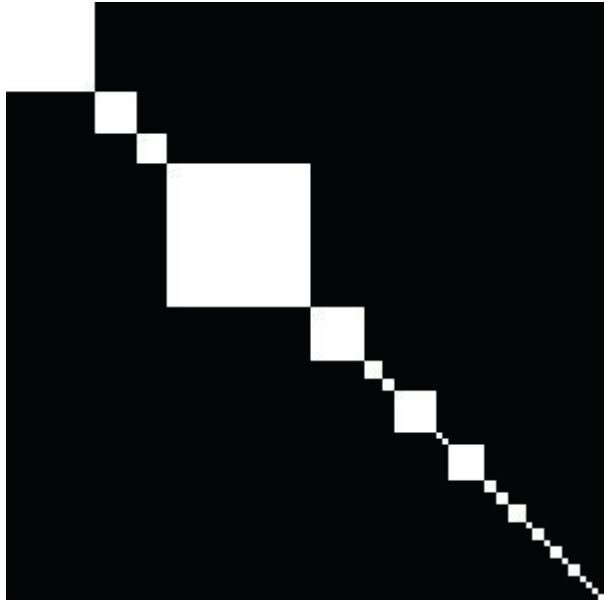}\\
\end{center}
\vspace{-0.4cm}
\caption{
Image matrix $\beta_{ij}$ at the 60,000th round from a random initial state, 
where $N$=100, $p$=0.7, $q$=1, and $\epsilon$=0.
}
\end{figure}
%%%%%%%%%%%%%%%%%%%%%%%%%%%%%%%%%%%%%%%%%%%%%%%%%%%%% 

%%%%%%%%%%%%%%%%%%%%%%%%%%%%%%%%%%%%%%%%%%%%%%%%%%%%%
\begin{figure}[tbhp]
\begin{center}
\includegraphics[width=90mm]{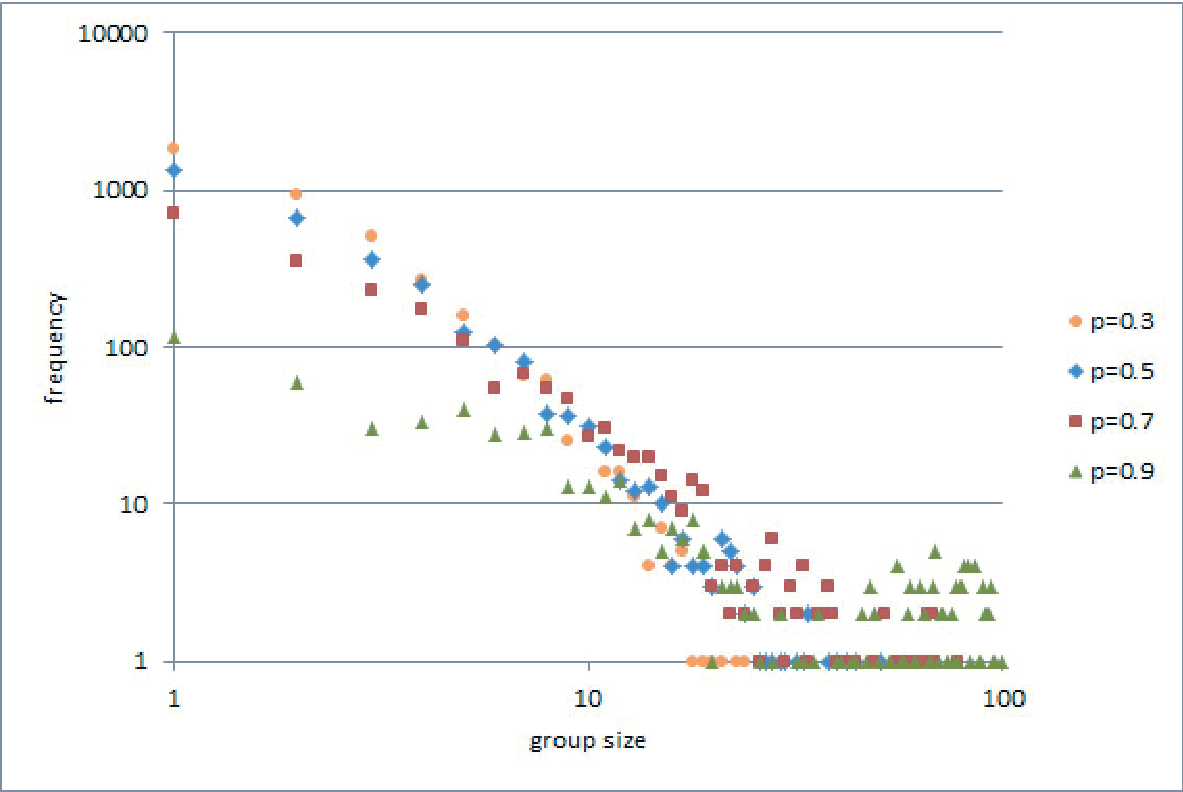}
\end{center}
\vspace{-0.4cm}
\caption{
Relations between group size and frequency 
for $p$=0.3, 0.5, 0.7, and 0.9, where $N$=100, $q$=1, 
and $\epsilon$=0. The frequency is a sum over 100 samples.
}
\end{figure}
%%%%%%%%%%%%%%%%%%%%%%%%%%%%%%%%%%%%%%%%%%%%%%%%%%%%% 

%%%%%%%%%%%%%%%%%%%%%%%%%%%%%%%%%%%%%%%%%%%%%%%%%%%%%
\begin{figure}[tbhp]
\begin{center}
\includegraphics[width=90mm]{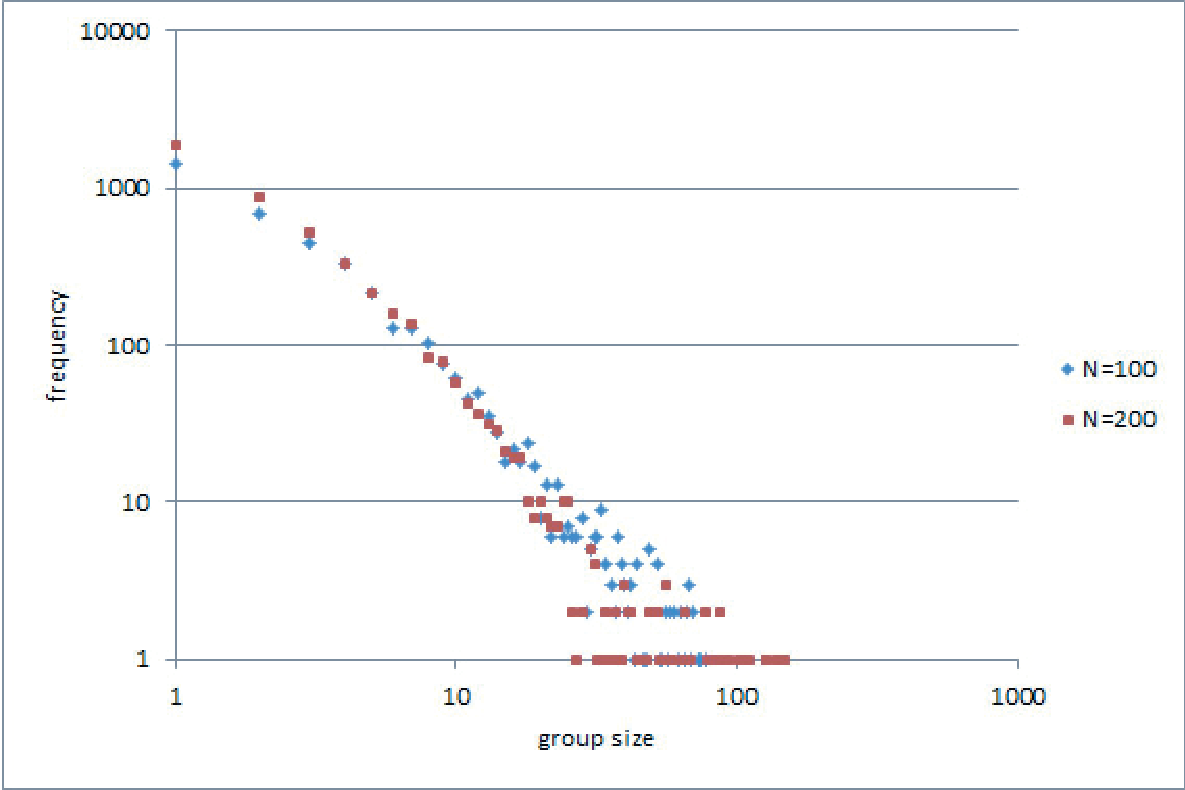}
\end{center}
\vspace{-0.4cm}
\caption{
Relations between group size and frequency 
for $N$=100 and 200, where $p$=0.7, $q$=1, and $\epsilon$=0. 
The frequency is a sum over 200 samples 
for $N$=100 and over 100 samples for $N$=200.
}
\end{figure}
%%%%%%%%%%%%%%%%%%%%%%%%%%%%%%%%%%%%%%%%%%%%%%%%%%%%% 

	Because $DEG1$ is only applicable to two exclusive groups,
 we introduce the following function for the degree of the exclusive groups 
for the JUDGING assessment rule ($DEG2$):
\begin{eqnarray}
DEG2 = \frac{\sum_{i=2}^{N}\sum_{j=1}^{i-1}(\beta_{ij}+1)(\beta_{ji}+1)/4}{\sum_{i=2}^{N}\sum_{j=1}^{i-1}(\beta_{ij}+1)/2}
\end{eqnarray}
$DEG2$ is one when many exclusive groups are perfectly formed, and zero when they are destroyed.@

\subsection{The case of $q<1$ and $\epsilon=0$}
In this case, many smaller exclusive groups are formed from a random
initial state (see Fig. 8), although $DEG2$ still remains one.
Once the formation of exclusive groups has been completed,
imperfection of information,
$1-q$, does not change the state(i.e. the state is stationary).
This behavior results in a decrease in good assessments.
Fig. 9 shows the fraction of good assessments as a function of $q$.

%%%%%%%%%%%%%%%%%%%%%%%%%%%%%%%%%%%%%%%%%%%%%%%%%%%%%
\begin{figure}[tbhp]
\begin{center}
\includegraphics{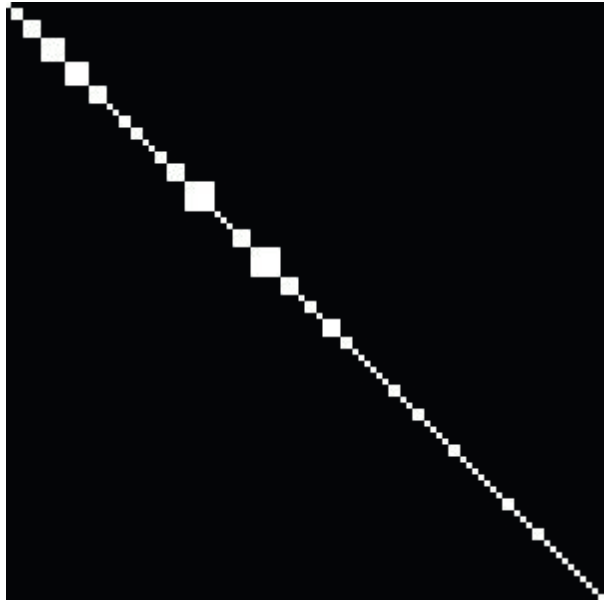}
\end{center}
\vspace{-0.4cm}
\caption{
Image matrix $\beta_{ij}$ at the 60,000th round from a random initial state, 
where $N$=100, $p$=0.7, $q$=0.7, and $\epsilon$=0.
}
\end{figure}
%%%%%%%%%%%%%%%%%%%%%%%%%%%%%%%%%%%%%%%%%%%%%%%%%%%%% 

%%%%%%%%%%%%%%%%%%%%%%%%%%%%%%%%%%%%%%%%%%%%%%%%%%%%%
\begin{figure}[tbhp]
\begin{center}
\includegraphics[width=90mm]{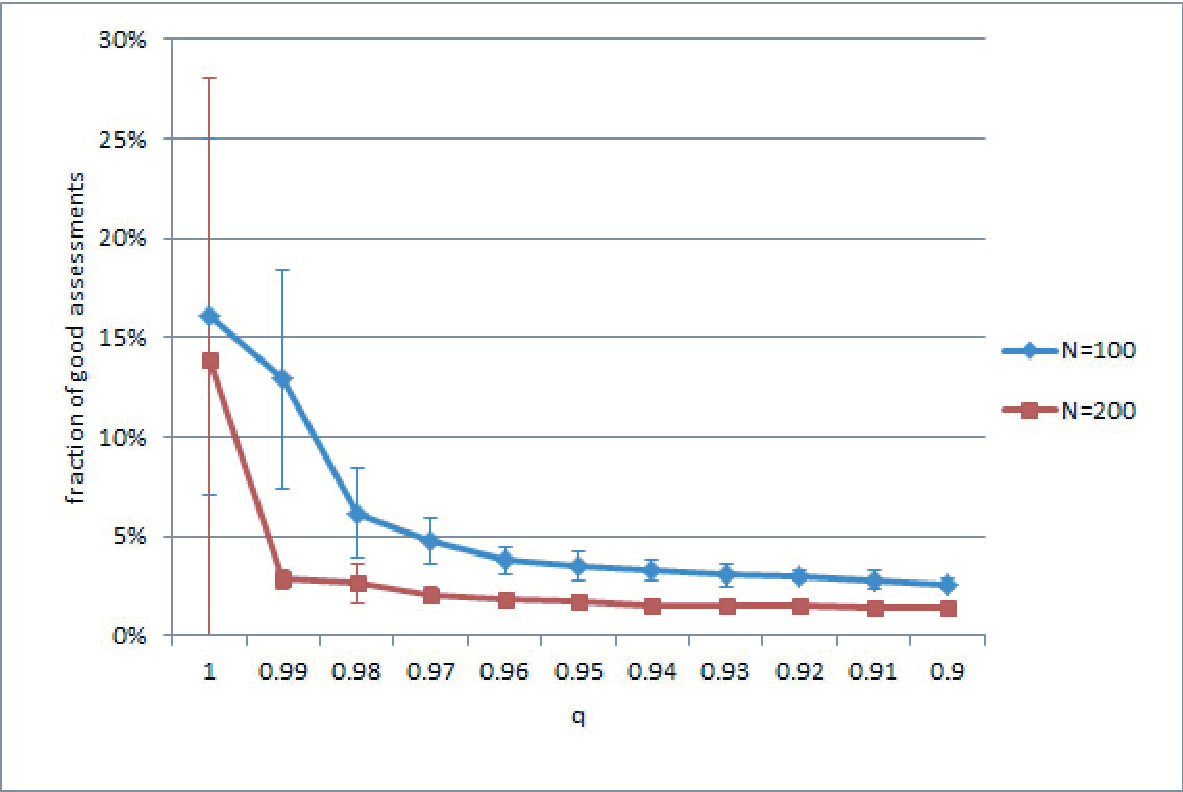}
\end{center}
\vspace{-0.4cm}
\caption{
Fraction of good assessments in the
image matrix $\beta_{ij}$ at the 60,000th round, starting with a random initial state, 
where $N$=100, $p$=0.7, $q$=0.7, and $\epsilon$=0. Error bars show the standard
deviation of 10 samples.
}
\end{figure}
%%%%%%%%%%%%%%%%%%%%%%%%%%%%%%%%%%%%%%%%%%%%%%%%%%%%% 

\subsection{The case of $q=1$ and $\epsilon>0$}
The degree of exclusive groups ($DEG2$) decreases almost linearly with $\epsilon$,
similar to the case of the KANDORI assessment rule.  
This does not necessarily mean the collapse of the group. 
As seen in Fig. 10, there are some white horizontal lines in addition to squares representing 
the exclusive groups. This indicates that there are players to 
assess the players in the different group as good: 
i.e., the exclusivity of the group formed in this case is not complete.

%%%%%%%%%%%%%%%%%%%%%%%%%%%%%%%%%%%%%%%%%%%%%%%%%%%%%
\begin{figure}[tbhp]
\begin{center}
\includegraphics{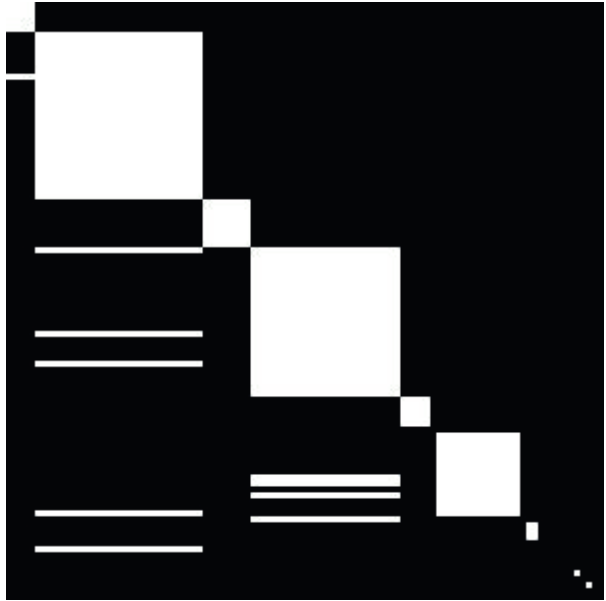}
\end{center}
\vspace{-0.4cm}
\caption{
Image matrix $\beta_{ij}$ at the 60,000th round starting with a random initial state, 
where $N$=100, $p$=0.9, $q$=1, and $\epsilon$=0.1.
}
\end{figure}
%%%%%%%%%%%%%%%%%%%%%%%%%%%%%%%%%%%%%%%%%%%%%%%%%%%%% 

\subsection{The case of $q<1$ and $\epsilon>0$}
The degree of the exclusive groups($DEG2$) decreases and groups are destroyed.
We also examine the stability against mutants with different assessment rules.
We consider the world of players with two or three different
assessment rules in Table 1, and compare the average total payoffs.
One example is shown in Fig. 11.
The results show that, with the exception of the STANDING mutants,
the JUDGING players have a larger average payoff than that of the other
mutants, although $DEG2$ decreases in the case of $q<1$ and $\epsilon>0$.

%%%%%%%%%%%%%%%%%%%%%%%%%%%%%%%%%%%%%%%%%%%%%%%%%%%%%
\begin{figure}[tbhp]
\begin{center}
\includegraphics[width=90mm]{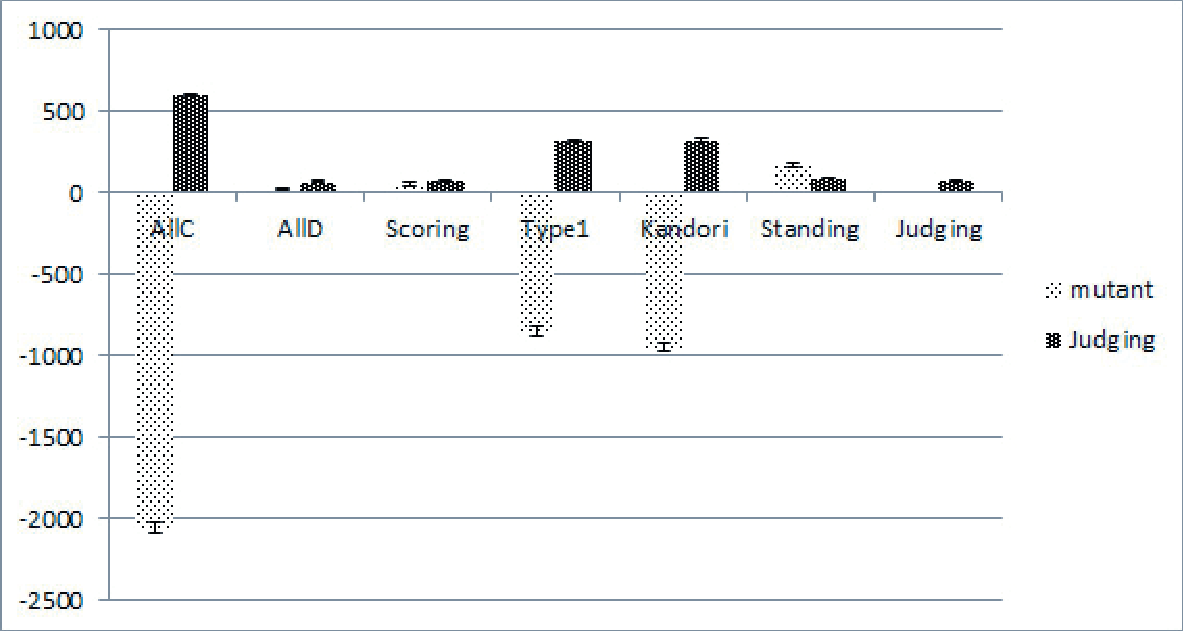}
\end{center}
\vspace{-0.4cm}
\caption{
Average total payoff at the 60,000th round in the world of players
with two assessment rules (90 JUDGING players and 10 mutants),
where $N$=100, $p$=0.7, $q$=0.9, $\epsilon$=0.1, $b$=10, and $c$=5.
Error bars show the standard deviation of 10 samples.
}
\end{figure}
%%%%%%%%%%%%%%%%%%%%%%%%%%%%%%%%%%%%%%%%%%%%%%%%%%%%% 

\section{Summary and Discussion}

	We summarize the results of this study as follows:
	
1) In contrast with the claims of Uchida\cite{Uchida}, 
imperfect information does not completely destroy two
exclusive groups in the KANDORI assessment rule.
For a small population size of $N$ and small imperfect
information $1-q$, two incomplete exclusive groups are formed.

2) Using the JUDGING assessment rule with $q=1$ and $\epsilon=0$, 
many exclusive groups
appear from a random initial state with a trust probability $p$,
where $0<p<1$.
 Distributions of the group sizes of exclusive groups show nearly
 power-law behaviors when $p$ is small and do not depend upon
 population size $N$. Incomplete information $1-q$ does not
decrease the degree of exclusive groups ($DEG2$), but makes their
group sizes smaller, resulting in an increased number
of bad assessments. 

3) If $b$ is sufficiently larger than $c$,
JUDGING players have larger average payoffs than 
mutants (other than STANDING mutants)
 with different assessment rules, as shown in Table 1, 
 even in the case where
$q<1$ and $\epsilon>0$.    

	The model investigated here has considerable limitations.
Introducing more realistic effects to the model may overcome
the negative points of the JUDGING players (namely smaller average
payoffs than those of the STANDING players and an increased number of
bad assessments due to the incomplete information). 
One of the limitations is that players have a very short memory:
they determine their opinions of another player according
to their last observation of that player's actions.
In real life, images are likely to be based on a longer memory.
In fact, we observe that introducing a longer memory to the model
makes the two exclusive groups of KANDORI players more robust under 
incompleteness of information\cite{shira}.

The other limitation is that the images are binary in this
model, whereas the moral world is not just black or white.
Tanabe et al. \cite{Tanabe} investigated the indirect reciprocity with
trinary reputations and found that this model allows
cooperation under the SCORING assessment rule under some mild conditions.
We may expect an increase in good assessments for the JUDGING players
under the imperfect information by changing the binary images into 
trinary ones.

% If you have acknowledgments, this puts in the proper section head.
\begin{acknowledgements}

We are thankful for the fruitful discussions with Professors F. Matsubara
and N. Suzuki. 
Part of the results in this research was obtained
 using supercomputing resources
at Cyberscience Center, Tohoku University.
% put your acknowledgments here.
\end{acknowledgements}

\appendix
\section{Sorting image matrices}
	We sort image matrices to make their structure easier to see. 
First, matrices are sorted based on the image of the first row. 
The columns are also sorted in the same order. At this time, a white square
 is formed in the upper left corner. In the case of the KANDORI assessment rule, 
a white square is also formed in the lower right corner, 
and sorting is completed. However, in the case of the JUDGING assessment rule, 
the lower right area has not been sorted. 
Therefore, the lower right area is sorted in the same way. 
This process is repeated until sorting becomes impossible. 
Finally, many white squares are formed on the diagonal of the image matrix. 
In other words, the players are divided into many exclusive groups.

% Create the reference section using BibTeX:
%%%%%%%%%%%%%%%%%%%%%%%\bibliography{basename of .bib file}

%

\end{document}